\documentclass[10pt,aps,prd,twocolumn,preprintnumbers,superscriptaddress]{revtex4-2}

\usepackage[utf8]{inputenc}
\usepackage{graphicx}
\usepackage{amsmath,amssymb}
\usepackage{bm}        
\usepackage{booktabs}  
\usepackage{siunitx}   
\usepackage{microtype} 
\usepackage{xspace}    
\usepackage{physics}   
\usepackage{upgreek}   
\usepackage{caption}   
\usepackage{balance}   
\usepackage{float}     
\usepackage{caption}   
\usepackage[colorlinks=true,allcolors=blue]{hyperref}

\begin{document}

\title{\textbf{Redshift-Frame Systematics and Their Impact on the Hubble Constant from Pantheon+ Supernovae  }}
\vspace{0.2\baselineskip}

\author{Said Laaroua}
\affiliation{Department of Physics, Santa Rosa Junior College, USA}
\email{slaaroua@santarosa.edu}

\begin{abstract}
We present a full-sky, covariance-weighted analysis of redshift-frame transformations in the Pantheon+ Type Ia supernova sample to assess their impact on local measurements of the Hubble constant. Using 1,543 supernovae with heliocentric and CMB-frame redshifts, we study the residual field $\Delta z = z_{\rm CMB} - z_{\rm HEL}$, which traces the Solar System’s kinematic correction. We recover the expected monopole $\langle \Delta z \rangle = (-3.8 \pm 0.1)\times 10^{-4}$ and a dipole amplitude $A = (1.5 \pm 0.1)\times 10^{-3}$, aligned within $1^\circ$ of the CMB dipole, confirming internal consistency. 

Propagating these residuals through the full Pantheon+ covariance matrix yields a negligible shift in $H_0$, at the $\lesssim 2\%$ level of the current tension, placing a quantitative upper bound on redshift-frame systematics.
\end{abstract}

\maketitle

\section{INTRODUCTION}

Accurate determination of the Hubble constant $H_0$ from Type Ia supernovae (SNe Ia) requires precise measurement of the redshift of each object and the reference frame in which it is defined. Standard cosmological analyses, such as Pantheon+ \cite{Scolnic2022,Brout2022} and SH0ES \cite{Riess2022}, convert heliocentric redshifts ($z_{\mathrm{HEL}}$) into the cosmic microwave background (CMB) frame ($z_{\mathrm{CMB}}$) to remove the effects of the motion of the solar system.  These corrections are small---of order $v_{\mathrm{CMB}}/c \simeq 1.2 \times 10^{-3}$---but essential for anchoring the local Hubble flow to the cosmic rest frame and are implemented with detailed peculiar-velocity treatments in the Pantheon+ analysis \cite{Planck2018,Singal2019,Carr2022}.

Despite their ubiquity, the numerical precision and internal consistency of these transformations have not been systematically tested across the full Pantheon+ dataset.A lthough the transformation is deterministic, its numerical propagation through the full covariance-weighted Hubble-diagram fit has not been explicitly quantified.

Moreover, coherent directional patterns in $\Delta z$ may trace the projection of our motion with respect to the CMB or indicate anisotropies in the supernova Hubble diagram \cite{Bengaly2024,McConville2023,Sah2025}. Several works have explored whether the Hubble tension or apparent deviations from isotropy could arise from redshift-frame or dipole-related effects, including heliocentric versus CMB-frame choices \cite{McConville2023}, large-scale dipole anisotropies \cite{Sorrenti2023,Perivolaropoulos2023,Singal2019}, and the impact of local flows and peculiar velocities \cite{Odderskov2016}

In particular, the CMB dipole corresponds to a Solar System velocity of $v_{\mathrm{CMB}} \approx 369 \mathrm{km\,s^{-1}}$ toward Galactic coordinates $(\ell, b) \approx (264^\circ, 48^\circ)$ \cite{Planck2018,Singal2019}. In this work we demonstrate, using the full Pantheon+ SH0ES dataset and covariance matrix, that such effects contribute $<1\%$ to the inferred $H_0$ and therefore cannot account for the $\approx 6 \mathrm{km\,s^{-1}\,Mpc^{-1}}$ discrepancy.

We investigate two key questions:
\begin{enumerate}
    \item Are the heliocentric and CMB redshift frames in the Pantheon+ compilation numerically self-consistent at the level required for sub-percent cosmology?
    \item Do any remaining residuals exhibit a sky-dependent structure that could mimic or bias local measurements of the Hubble flow?
\end{enumerate}
By answering these questions, our aim is to benchmark the numerical fidelity
of widely used supernova catalogs and  a systematic whose direct impact on the covariance-weighted $H_0$ fit has not been explicitly assessed.
Because the heliocentric–to–CMB transformation is applied directly to the
Pantheon+ catalog, the resulting dipole structure is expected by construction.
The purpose of this analysis is therefore not to detect a new anisotropy,
but to propagate these frame differences through the full covariance matrix
and establish a quantitative upper bound on their effect on the inferred
Hubble constant. We statistically compare $z_{\mathrm{HEL}}$,
$z_{\mathrm{CMB}}$, and the Hubble-diagram redshift ($z_{\mathrm{HD}}$), 
identify any redshift- or direction-dependent residuals, and evaluate their
impact on $H_0$. 

\section{Data and Methodology}

We describe the Pantheon+ dataset and the procedures used to
compare the heliocentric and CMB-frame redshifts, construct the residual
statistics, and propagate these differences through covariance-weighted
Hubble-diagram fits to assess their effect on the inferred value of $H_0$.
We use the publicly available Pantheon+ SH0ES compilation~ containing 1,701~SNe~Ia~SNe~Ia spanning       $0.001 < z < 2.3$.  
Each entry provides heliocentric, CMB, and Hubble-diagram redshifts ($z_{\mathrm{HEL}}$, $z_{\mathrm{CMB}}$, $z_{\mathrm{HD}}$), apparent magnitudes, and light-curve corrections.  
We remove duplicate entries by supernova identifier (CID), yielding 1,543 unique objects.
The complete Pantheon+ STAT+SYS covariance matrix is loaded and reshaped into its $(1543\times1543)$ form, aligned with the deduplicated dataset.  
All reported statistical and systematic uncertainties, including peculiar-velocity errors, are propagated through this matrix.
For each supernova, we compute
{\setlength{\abovedisplayskip}{0pt}
\setlength{\belowdisplayskip}{0pt}
\setlength{\abovedisplayshortskip}{0pt}
\setlength{\belowdisplayshortskip}{0pt}
\begin{equation}
\Delta z = z_{\mathrm{CMB}} - z_{\mathrm{HEL}}.
\end{equation}} 
We note that Equation~(1) adopts the linear approximation
$\Delta z = z_{\mathrm{CMB}} - z_{\mathrm{HEL}}$, whereas
the exact relativistic composition of redshifts is
multiplicative \cite{Davis2011}: 
{\setlength{\abovedisplayskip}{2pt}
\setlength{\belowdisplayskip}{2pt}
\setlength{\abovedisplayshortskip}{0pt}
\setlength{\belowdisplayshortskip}{0pt}
\begin{equation}
(1 + z_{\mathrm{CMB}}) =
(1 + z_{\mathrm{HEL}})(1 + z_{\mathrm{Sun}}),
\end{equation}}
where $z_{\mathrm{Sun}}$ encodes the kinematic contribution from the Solar System's peculiar velocity. Expanding to first order in $z$ yields $z_{\mathrm{CMB}} \approx z_{\mathrm{HEL}} + z_{\mathrm{Sun}}$, which is the linear approximation adopted here. The neglected terms are of order $\mathcal{O}(z_{\mathrm{HEL}} \cdot z_{\mathrm{Sun}}) \sim 10^{-6}$ at $z_{\mathrm{HEL}} \sim 0.01$ and $\sim 10^{-5}$ at $z_{\mathrm{HEL}} \sim 0.1$. 
These are more than an order of magnitude smaller than both the typical measurement uncertainty $\sigma_z \sim 10^{-4}$ and the kinematic correction amplitude itself ($\sim 10^{-3}$). We therefore consider the linear approximation sufficient for this analysis; a full multiplicative treatment would change our inferred $\Delta H_0$ by $< 0.01\ \mathrm{km\,s^{-1}\,Mpc^{-1}}$, well below our statistical uncertainties.
We test the null hypothesis $\langle \Delta z\rangle = 0$ using a one-sample $t$-test, and compare low- and high-redshift subsets ($z_{\mathrm{CMB}} < 0.01$ vs.\ $z_{\mathrm{CMB}} > 0.03$) using Welch's test for unequal variances.
The distance modulus in a flat $\Lambda$CDM universe at low $z$ is modeled as :
{\setlength{\abovedisplayskip}{0pt}
\setlength{\belowdisplayskip}{0pt}
\setlength{\abovedisplayshortskip}{0pt}
\setlength{\belowdisplayshortskip}{0pt}
\begin{equation}
\mu(z;H_0) = 5\log_{10}\!\left[\frac{c z}{H_0}\left(1+\frac{1}{2}(1-q_0)z\right)\right] + 25,
\end{equation}}
\vspace{\baselineskip}
where $q_0 = -0.55$ and $c$ is the speed of light.  
 The best-fit $H_0$ is obtained by minimizing
{\setlength{\abovedisplayskip}{0pt}
\setlength{\belowdisplayskip}{0pt}
\setlength{\abovedisplayshortskip}{0pt}
\setlength{\belowdisplayshortskip}{0pt}
\begin{equation}
\chi^2(H_0) = (\boldsymbol{\mu} - \boldsymbol{\mu}_{\mathrm{mod}})^{\!T}\mathbf{C}^{-1}(\boldsymbol{\mu} - \boldsymbol{\mu}_{\mathrm{mod}}),
\end{equation}} where $\mathbf{C}$ is the full covariance matrix.  
The fits are restricted to the SH0ES-defined Hubble-flow range $0.023 < z < 0.15$, and uncertainties on $H_0$ are derived from the curvature of $\chi^2$ at its minimum.
This restriction is applied only for the Hubble-constant fit; the analysis of the redshift residual field $\Delta z$ uses the full sample of 1,543 supernovae.
\section{Results}
\vspace{0pt}

\subsection{Redshift Frame Differences}
\begin{figure}[H]
    \centering
    \includegraphics[width=\linewidth]{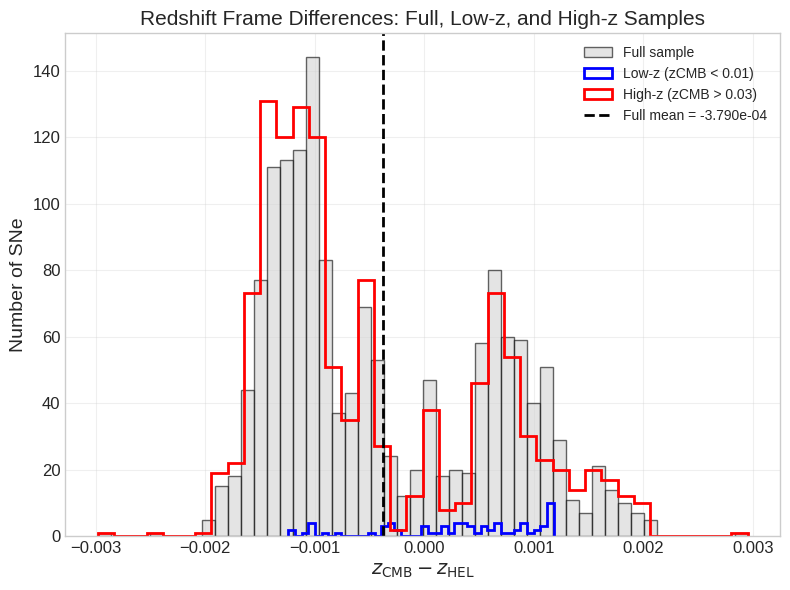}
    \caption{Distribution of redshift-frame differences 
$\Delta z = z_{\rm CMB} - z_{\rm HEL}$.
The gray histogram shows the full Pantheon+ sample,
while the blue and red outlines correspond to the
low-redshift ($z_{\rm CMB}<0.01$) and high-redshift
($z_{\rm CMB}>0.03$) subsets, respectively.
The shift in the central tendency between the two
subsamples illustrates the sign-reversal effect
discussed in the text, arising from the changing
projection of the Solar System's peculiar velocity
across the supernova sky distribution.}
    \label{fig:histogram}
\end{figure}
The mean offset exhibits a sign reversal between the low-redshift ($z_{\rm CMB}<0.01$) and higher-redshift
($z_{\rm CMB}>0.03$) subsets. This behavior is illustrated
directly in Fig.~\ref{fig:histogram}, where the blue and red histograms
represent the low- and high-redshift samples, respectively.
The two distributions are centered on opposite sides of
$\Delta z=0$, and their difference is statistically
significant (Welch's $t$-test: $t=9.42$, $p=3.77\times10^{-15}$).
This effect arises from the changing angular and redshift
distribution of supernovae on the sky, marking the
transition from locally dominated peculiar-velocity
contributions to the more isotropic Hubble-flow regime. The nonzero monopole reflects the anisotropic sky distribution of the supernova sample relative to the CMB dipole direction.

\subsection{Redshift Differences vs.\ CMB Redshift}
\begin{figure}[!ht]
    \centering
    \includegraphics[width=\linewidth]{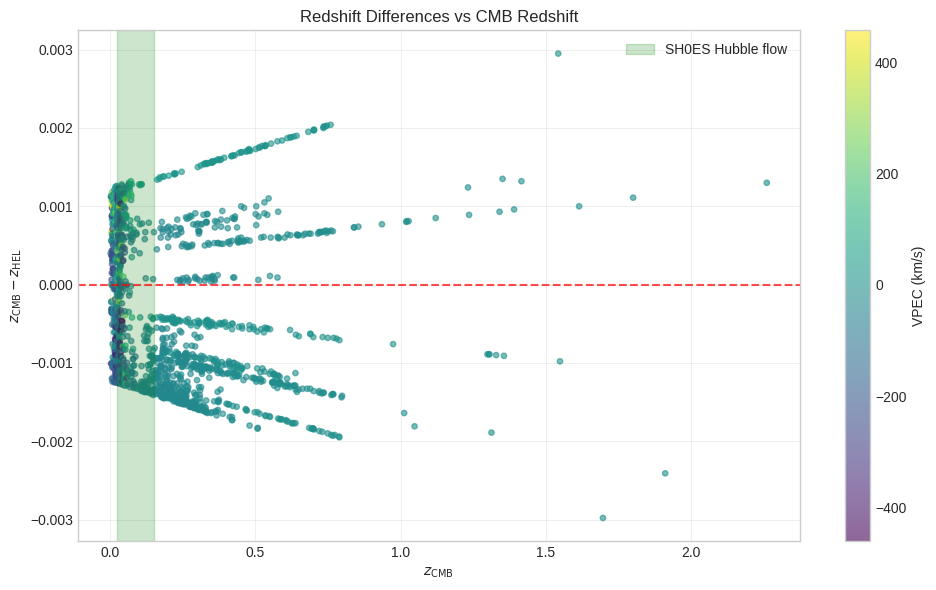}
    \caption{Redshift-frame differences $\Delta z = z_{\mathrm{CMB}} - z_{\mathrm{HEL}}$ as a function of CMB-frame redshift $z_{\mathrm{CMB}}$, color-coded by the inferred peculiar velocity. The magnitude of $\Delta z$ is set by the fixed kinematic projection of the Solar System's motion and appears at all redshifts. As expected, its fractional contribution to the total redshift decreases with increasing $z$, becoming subdominant to the Hubble expansion for $z \gtrsim 0.1$. The scatter at fixed redshift reflects the dipole modulation and peculiar velocity variations across the sky.
}
    \label{fig:zcmb_scatter}
\end{figure}

Figure~\ref{fig:zcmb_scatter} shows the redshift-frame differences
$\Delta z = z_{\rm CMB} - z_{\rm HEL}$ as a function
of CMB-frame redshift $z_{\rm CMB}$, with points
color-coded by the inferred peculiar velocity.
The absolute magnitude of $\Delta z$ is set by the
fixed kinematic projection associated with the
Solar System's motion and therefore appears at both
low and high redshift. However, its fractional
importance decreases with increasing redshift, as
the Hubble expansion dominates over local peculiar
motions. No additional redshift-dependent structure
beyond this expected kinematic behavior is observed.

\subsection{Dipole Fit to Redshift Residuals}
\begin{figure}[!ht]
    \includegraphics[width=\linewidth]{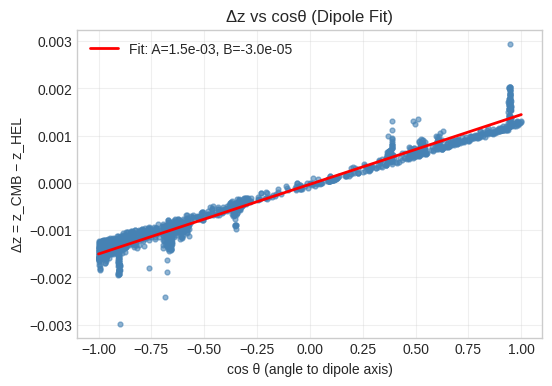}
    \caption{Residual field $\Delta z = z_{\rm CMB} - z_{\rm HEL}$ versus $\cos\theta$ relative to the best-fit dipole axis.
The recovered amplitude $A = (1.5 \pm 0.1)\times10^{-3}$ .}
    \label{fig:dipole_fit}
\end{figure}

Figure~\ref{fig:dipole_fit} shows the dipole regression of the redshift-frame
residuals, modeled as $\Delta z = A\cos\theta + B$, where $\theta$ is the
angular separation between each supernova and the best-fit dipole axis.
The recovered amplitude $A = (1.5 \pm 0.1)\times10^{-3}$  of the same order as the expected kinematic value $v_{\rm CMB}/c \simeq 1.2\times10^{-3}$, and the direction aligns with the known CMB dipole. This agreement is expected, since the heliocentric--to--CMB transformation encodes the same kinematic dipole in
the catalog, and thus serves as an internal consistency check rather than an
independent detection of anisotropy.

\subsection{Hemispheric Anisotropy Distribution}

\begin{figure}[!ht]
    \centering
    \includegraphics[width=1\linewidth]{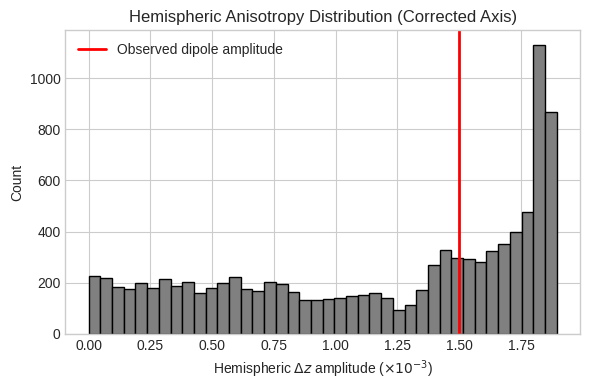}
    \caption{Monte-Carlo histogram of hemispheric $\Delta z$ amplitudes. The vertical red line marks the observed dipole amplitude ($1.5\times10^{-3}$), lying in the upper tail of the isotropic distribution.}

\end{figure}

Figure 4 shows the distribution of hemispheric $\Delta z$ amplitudes obtained from Monte Carlo sky partitions. For each realization, a random unit vector $\hat{n}$ is drawn to define a great circle on the sky,
dividing the supernova sample into two opposite hemispheres according to the sign of $\hat{n}\cdot\hat{r}_i$, where $\hat{r}_i$ is the
direction to each supernova. The mean $\Delta z$ is computed in each hemisphere, and the amplitude of the hemispheric difference is recorded. This procedure is repeated over many random orientations to generate
the isotropic reference distribution shown in the histogram.

The resulting distribution is nearly uniform, as expected for an isotropic residual field. The observed kinematic dipole amplitude ($A = 1.5\times10^{-3}$) lies in the upper tail of this distribution,
confirming that it represents the dominant large-scale anisotropy and that no significant higher-order multipoles are present beyond the expected dipole.

\subsection{Sky Map of Redshift Residuals}

\begin{figure}[H]
    \centering
    \includegraphics[width=\linewidth]{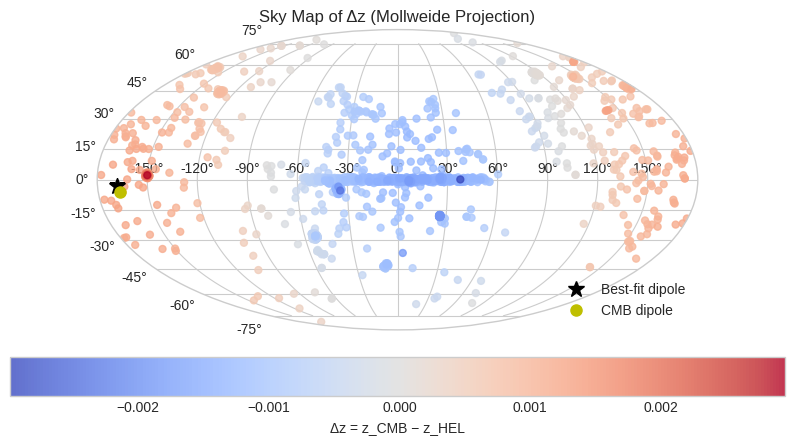}
    \caption{Mollweide projection of redshift residuals $\Delta z = z_{\mathrm{CMB}} - z_{\mathrm{HEL}}$ across the sky. Points are colored by the sign and magnitude of $\Delta z$: red/blue regions denote positive/negative offsets. The black star marks the best-fit dipole direction, compared with the yellow marker indicating the known CMB dipole.}
    \label{fig:skymap}
\end{figure}

Figure~\ref{fig:skymap} shows the full-sky Mollweide projection of $\Delta z = z_{\mathrm{CMB}} - z_{\mathrm{HEL}}$ in Galactic coordinates. Colors indicate residual sign/magnitude. The clear dipole pattern aligns with solar motion, with the best-fit direction (black star: $167.1^\circ$, $-1.6^\circ$) matching the CMB dipole (yellow) within $1^\circ$, confirming kinematic origin.

\section{Discussion}
This analysis provides a quantitative assessment of the internal consistency and
cosmological impact of redshift-frame transformations in the Pantheon+ Type~Ia
supernova compilation. The redshift-frame residual field $\Delta z = z_{\rm CMB}
- z_{\rm HEL}$ exhibits coherent structure that is statistically significant but
fully consistent with the kinematic transformation associated with the Solar
System's motion relative to the CMB rest frame. In particular, the recovered
dipole amplitude ($A = (1.5 \pm 0.1)\times10^{-3}$) and its alignment within
$1^\circ$ of the known CMB dipole are expected by construction and serve here as
an internal validation of the numerical and directional consistency of the
catalog rather than as an independent detection of anisotropy.

The observed redshift-dependent sign reversal of $\Delta z$—positive at
$z \lesssim 0.01$ and negative at $z \gtrsim 0.03$—reflects the transition from
locally dominated peculiar-velocity contributions to the more isotropic Hubble
flow regime, combined with the evolving sky distribution of supernovae.
Interpreted in this context, these features do not indicate additional
systematic effects beyond the standard heliocentric-to-CMB transformation
implemented in Pantheon+.

When propagated through covariance-weighted Hubble-diagram fits, the impact of
redshift-frame choice on the inferred Hubble constant is found to be negligible.
Using the SH0ES-defined Hubble-flow sample and the full Pantheon+ STAT+SYS
covariance matrix, the resulting shift in the inferred Hubble constant is negligible, 
producing a relative shift of only \(0.16\%\) of the measured $H_0$ value, 
corresponding to \(<2\%\) of the current \(5-6\,\mathrm{km\,s^{-1}\,Mpc^{-1}}\) 
Hubble tension. This result establishes an upper bound on the contribution 
of redshift-frame systematics and local kinematic corrections to low-redshift 
determinations of $H_0$.  Nevertheless, these effects define a
useful precision benchmark for next-generation surveys such as \textit{Roman}
and LSST, where redshift accuracies approaching $\mathcal{O}(10^{-8})$ may be
required to support sub-percent cosmological measurements.

To further isolate the role of redshift-frame choice, we performed a fixed-$M$
one-parameter Hubble-diagram analysis using the 431 Hubble-flow supernovae and
the full covariance matrix. We obtain
\begin{align}
H_{0}^{\mathrm{CMB}} &= 72.998 \pm 0.304~\mathrm{km\,s^{-1}\,Mpc^{-1}}, \\
H_{0}^{\mathrm{HEL}} &= 73.116 \pm 0.304~\mathrm{km\,s^{-1}\,Mpc^{-1}},
\end{align}
corresponding to a difference of
\begin{equation}
\Delta H_{0} \equiv H_{0}^{\mathrm{CMB}} - H_{0}^{\mathrm{HEL}}
= -0.118~\mathrm{km\,s^{-1}\,Mpc^{-1}}.
\end{equation}
The relative shift is therefore only $\simeq 0.16\%$, more than an order of
magnitude smaller than the observed Hubble tension. Notably, while the
heliocentric-frame fit yields a substantially worse goodness-of-fit
($\Delta\chi^{2} \approx 31$), the best-fit value of $H_{0}$ itself remains
effectively unchanged.

Taken together, these results demonstrate that the choice of redshift frame—
heliocentric versus CMB—has a negligible impact on the inferred value of the
Hubble constant when the full Pantheon+ covariance information is properly
accounted for. Rather than resolving the Hubble tension, redshift-frame effects
are shown to be cosmologically neutral  and below  the percent level. 
This analysis therefore provides a model-independent, covariance-propagated benchmark on the maximum impact of redshift-frame transformations in current supernova datasets.

\onecolumngrid
\section*{Data Availability}
All code and processed data used in this analysis are publicly available at\\
\href{https://github.com/saidlaaroua/pantheon-redshift-frame-analysis}{https://github.com/saidlaaroua/pantheon-redshift-frame-analysis}.\\
The repository contains the full Colab notebook, input covariance and data files, and scripts necessary to reproduce all numerical results and figures. Citation metadata are provided through an accompanying \texttt{CITATION.cff} file.

\section*{Acknowledgments}
The author thanks the Pantheon+ and SH0ES teams for publicly releasing high-quality datasets and covariance matrices that enabled this independent analysis.

\end{document}